\documentclass[aps,prl,superscriptaddress,twocolumn,nopacs,letter,rpreprint]{revtex4-1}
\usepackage{graphicx,bm,times}
\usepackage{amsmath}
\usepackage{amsfonts}
\usepackage{amssymb}
\usepackage{color}
\usepackage{hyperref}
\hypersetup{
	colorlinks = true,
	allcolors = {blue}
}
\setcitestyle{super}
\usepackage[utf8]{inputenc}
\usepackage[T1]{fontenc}
\usepackage{graphicx}
\usepackage{amsmath,amssymb}
\begin{document}
	
\title{Electronically driven spin-reorientation transition of the correlated polar metal Ca$_3$Ru$_2$O$_7$}
	
\author{I.~Markovi{\'c}}
\affiliation {SUPA, School of Physics and Astronomy, University of St Andrews, St Andrews KY16 9SS, United Kingdom}
\affiliation{Max Planck Institute for Chemical Physics of Solids, N\"{o}thnitzer Strasse 40, 01187 Dresden, Germany}

\author{M.~D.~Watson}
\author{O.~J.~Clark}
\author{F.~Mazzola}
\affiliation {SUPA, School of Physics and Astronomy, University of St Andrews, St Andrews KY16 9SS, United Kingdom}

\author{E.~Abarca Morales}
\affiliation{Max Planck Institute for Chemical Physics of Solids, N\"{o}thnitzer Strasse 40, 01187 Dresden, Germany}
\affiliation {SUPA, School of Physics and Astronomy, University of St Andrews, St Andrews KY16 9SS, United Kingdom}

\author{C.~A.~Hooley}
\affiliation {SUPA, School of Physics and Astronomy, University of St Andrews, St Andrews KY16 9SS, United Kingdom}

\author{H.~Rosner}
\affiliation{Max Planck Institute for Chemical Physics of Solids, N\"{o}thnitzer Strasse 40, 01187 Dresden, Germany}

\author{C.~M.~Polley}
\author{T.~Balasubramanian}
\affiliation{MAX IV Laboratory, Lund University, P. O. Box 118, 221 00 Lund, Sweden}

\author{S.~Mukherjee}
\affiliation {Diamond Light Source, Harwell Campus, Didcot, OX11 0DE, United Kingdom}

\author{N.~Kikugawa}
\affiliation{National Institute for Materials Science, Tsukuba, Ibaraki 305-0003, Japan}

\author{D.~A.~Sokolov}
\affiliation{Max Planck Institute for Chemical Physics of Solids, N\"{o}thnitzer Strasse 40, 01187 Dresden, Germany}

\author{A.~P.~Mackenzie}
\affiliation{Max Planck Institute for Chemical Physics of Solids, N\"{o}thnitzer Strasse 40, 01187 Dresden, Germany}
\affiliation {SUPA, School of Physics and Astronomy, University of St Andrews, St Andrews KY16 9SS, United Kingdom}
	
\author{P.~D.~C.~King}
\email{philip.king@st-andrews.ac.uk}
\affiliation {SUPA, School of Physics and Astronomy, University of St Andrews, St Andrews KY16 9SS, United Kingdom}

\date{\today}
\maketitle

\textbf{Polar distortions in solids give rise to the well-known functionality of switchable macroscopic polarisation in ferroelectrics~\cite{rabe_modern_2007,scott_applications_2007} and, when combined with strong spin-orbit coupling, can mediate giant spin splittings of electronic states~\cite{di_sante_electric_2013,picozzi_ferroelectric_2014}. While typically found in insulators, ferroelectric-like distortions can remain robust against increasing itineracy, giving rise to so-called ``polar metals''~\cite{anderson_symmetry_1965,shi_ferroelectric-like_2013,puggioni_designing_2014,benedek_ferroelectric_2016,kim_polar_2016,laurita_evidence_2019}. Here, we investigate the temperature-dependent electronic structure of Ca$_3$Ru$_2$O$_7$, a correlated oxide metal in which octahedral tilts and rotations combine to mediate pronounced polar distortions~\cite{yoshida_crystal_2005,lei_observation_2018}. Our angle-resolved photoemission measurements reveal the destruction of a large hole-like Fermi surface upon cooling through a coupled structural and spin-reorientation transition at 48~K, accompanied by a sudden onset of quasiparticle coherence. We demonstrate how these result from band hybridisation mediated by a hidden Rashba-type spin-orbit coupling. This is enabled by the bulk structural distortions and unlocked when the spin reorients perpendicular to the local symmetry-breaking potential at the Ru sites. We argue that the electronic energy gain associated with the band hybridisation is actually the key driver for the phase transition, reflecting a delicate interplay between spin-orbit coupling and strong electronic correlations, and revealing a new route to control magnetic ordering in solids.}     

\begin{figure*}
	\centering
	\includegraphics[width=\textwidth]{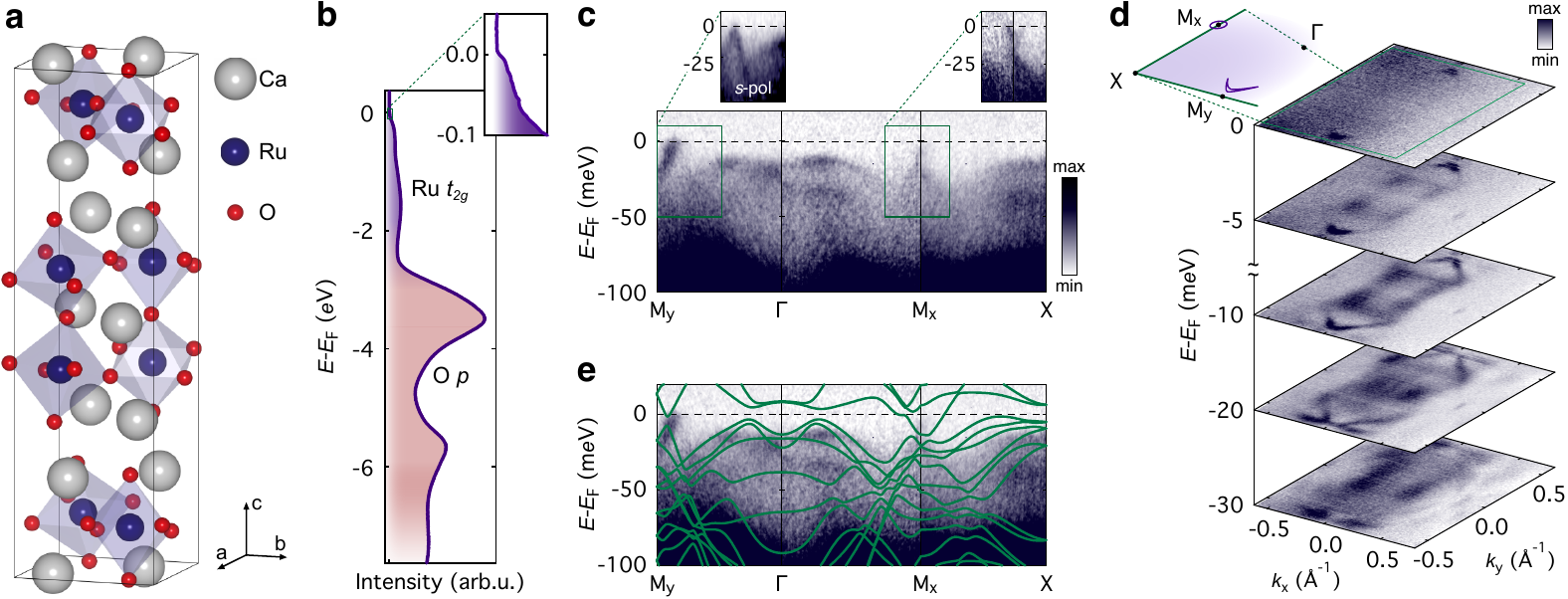}
	\caption{\textbf{Low-energy electronic structure of Ca$_3$Ru$_2$O$_7$.} (a) Crystal structure of Ca$_3$Ru$_2$O$_7$. (b) Overview electronic structure measured at $T=6$~K. A broad bandwidth associated with spectral weight from the Ru $t_{2g}$ states is observed, while the spectral weight becomes very small close to the Fermi level. (c) Despite this, sharp quasiparticle-like states are observed at low energy. Close to M$_y$, a hole band just intersects the Fermi level, with both branches of the dispersion more clearly visible for measurements performed using $s$-polarised light (left inset). At M$_x$, an electron-like band crosses $E_\mathrm{F}$, shown with enhanced contrast in the inset. (d) Constant energy contours measured at the Fermi level ($E_\mathrm{F}\pm2$~meV) and at finite binding energy. The Fermi surface is composed of tiny electron pockets located at the high-symmetry M$_x$ point, and small boomerang-shaped hole Fermi surfaces located away from the Brillouin zone boundary close to M$_y$, as shown schematically inset. Additional states with `M-shaped' dispersion are observed at slightly higher binding energy ((c), starting from $E-E_\mathrm{F}\approx10$~m{\em e}V), which are also visible in constant energy contours below the Fermi level. (e) While the large unit cell and magnetic ordering give rise to a multitude of bands, we find that the general structure and form of the near-$E_{\mathrm{F}}$ states are well captured by ferromagnetic density-functional theory calculations (see Methods) when scaling the energy by a factor of 0.15. }
	\label{fig1}
\end{figure*}

Ca$_3$Ru$_2$O$_7$ is the bilayer member of the Ca$_{n+1}$Ru$_n$O$_{3n+1}$ Ruddlesden-Popper series. The small ionic size of Ca induces large coupled rotations and tilts of the RuO$_6$ octahedra that make up the perovskite-like building blocks of this structure, generating a non-centrosymmetric crystal structure (space group \#36: {\em Bb2$_1$m}, see Fig.~\ref{fig1}(a))~\cite{yoshida_crystal_2005,lei_observation_2018}. A symmetry-allowed trilinear coupling between the two non-polar octahedral tilt ($X^-_3$) and rotation ($X^+_2$) modes and a polar lattice mode ($\Gamma^-_5$) further mediates polar distortions, just as in Ca$_3$Ti$_2$O$_7$ and Ca$_3$Mn$_2$O$_7$ which are part of the well-known class of improper ferroelectrics~\cite{bousquet_improper_2008,benedek_hybrid_2011,nowadnick_domains_2016}. Unlike these sister compounds, however, Ca$_3$Ru$_2$O$_7$ is not an insulator. Its in-plane resistivity decreases upon cooling from room temperature, albeit with a linear temperature dependence indicative of a so-called ``bad metal'' state~\cite{yoshida_quasi-two-dimensional_2004,kikugawa_ca3ru2o7:_2010}.  

At $T_\mathrm{N}=56$~K, the system undergoes a N{\'e}el ordering transition, where the spins align ferromagnetically within each bilayer, oriented along the ${a}$-axis, and are antiferromagnetically coupled between bilayers~\cite{bohnenbuck_magnetic_2008,bao_spin_2008}. At a second phase transition at $T_\mathrm{S}=48$~K, the spins reorient to lie parallel to the in-plane ${b}$-axis~\cite{bohnenbuck_magnetic_2008,bao_spin_2008,yoshida_crystal_2005}. Simultaneously, an iso-structural transition leads to a squashing of the unit cell along the $c$-direction~\cite{yoshida_crystal_2005}. The resistivity exhibits a sudden jump on cooling through $T_\mathrm{S}$, but, although its absolute value remains relatively high ($\rho_{ab}(5~\mathrm{K})\approx{50}$~$\mu\Omega$cm), the in-plane resistivity recovers a metallic temperature-dependence to low temperature~\cite{yoshida_quasi-two-dimensional_2004,kikugawa_ca3ru2o7:_2010}. A rich non-collinear magnetic texture has been observed under the application of magnetic fields~\cite{sokolov_metamagnetic_2019}, pointing to an important role of spin-orbit coupling combined with the non-centrosymmetric crystal structure, while strong electronic correlations are expected to also play an important role in shaping the electronic and magnetic properties of this system (the single-layered sister compound Ca$_2$RuO$_4$ is a Mott insulator~\cite{nakatsuji_ca_1997}).

Ca$_3$Ru$_2$O$_7$ thus stands as a particularly rich example of a correlated polar metal. Gaining a comprehensive understanding of its transport, magnetic, and electronic properties has, however, proved elusive to date. Here, we study its temperature-dependent electronic structure by angle-resolved photoemission (ARPES, see Methods). Our low-temperature measurements are shown in Fig.~\ref{fig1}. Consistent with Refs.~\cite{baumberger_nested_2006,horio_electron-driven_2019}, we find a significant spectral weight in the valence bands associated with the Ru $t_{2g}$ orbitals, but almost vanishing spectral weight at the Fermi level (Fig.~\ref{fig1}(b)). Nonetheless, sharp features are still evident on low energy-scales ($\lesssim50$~m{\em e}V, Fig.~\ref{fig1}(c)), indicative of well-defined Fermi liquid-like quasiparticles, albeit with very low quasiparticle residue.  

Our measured electronic structure exhibits a pronounced two-fold symmetry throughout the Brillouin zone. At the M$_x$ point, an electron band intersects the Fermi level, giving rise to a small M$_x$-centred electron pocket (just visible in Fig.~\ref{fig1}(d)). This derives from a band whose occupied bandwidth is only $\lesssim15$~m{\em e}V, immediately below which another band disperses downwards to higher binding energy (see right inset in Fig.~\ref{fig1}(c)). The electronic structure is markedly different along the $\Gamma$-M$_y$ direction. A dispersive state is evident intersecting the Fermi level away from the Brillouin zone boundary (Fig.~\ref{fig1}(c)). An extremely weak feature is also visible with approximately the same $k_{\mathrm{F}}$ but opposite Fermi velocity (confirmed using measurements with different light polarisation, see left inset in Fig.~\ref{fig1}(c)), indicating that this is the top of a $\Lambda$-shaped band which barely grazes the Fermi level. Fermi surface measurements (Fig.~\ref{fig1}(d)) show how this disperses along the perpendicular in-plane direction to form a very narrow boomerang-shaped hole-like Fermi surface, centred along the $\Gamma$-M$_y$ line but displaced away from the Brillouin zone boundary. 

We thus assign the ground state of Ca$_3$Ru$_2$O$_7$ to be a low carrier-density compensated semi-metal, in agreement with the small Fermi pockets found previously by de Haas van Alphen studies~\cite{kikugawa_ca3ru2o7:_2010}. An additional set of sharp and rather flat states are visible in Fig.~\ref{fig1}(c,d) closer to the Brillouin zone centre, with their band maxima $\approx\!10$~m{\em e}V below the Fermi level~\cite{baumberger_nested_2006}. Measurements using different light polarisations (Supplementary Fig.~S1) indicate that there are at least 3 distinct states here, pointing to a rich multi-band near-$E_\mathrm{F}$ electronic structure. To further explore this, we show in Fig.~\ref{fig1}(e) density-functional theory (DFT) calculations of the low-temperature electronic structure (see Methods), renormalised in energy by a factor of $\approx\!7$. 

The strong bandwidth renormalisation needed to achieve a reasonable agreement with the measured low-energy electronic structure indicates that Ca$_3$Ru$_2$O$_7$ is a highly correlated metal, consistent with its low quasiparticle residues. In such a complex multi-band system as this, momentum- and orbital-dependent self energies may generically be expected~\cite{baumberger_nested_2006}, and a simple bandwidth scaling cannot be expected to capture in detail the full influence of many-body interactions on the electronic structure. Indeed, the experimental Fermi velocities of the boomerang-shaped states crossing $E_{\mathrm{F}}$ are only renormalised by a factor of $\approx\!4$ as compared to corresponding features in the DFT, while the flat bands appear to require significantly higher renormalisations. The results shown here thus motivate future study of interaction effects in Ca$_3$Ru$_2$O$_7$ by state-of-the-art correlated electronic structure calculations, of the form that have recently proved extremely successful in describing the single-layer Sr-based sister compound~\cite{tamai_high-resolution_2019,acharya_evening_2019}. Nonetheless, we note that a global bandwidth scaling of the calculated DFT still does a remarkably good job in reproducing the key experimental band structure features observed here, including the M$_x$-centred electron pocket, the hole-like $\Lambda$-band offset from the M$_y$ point (located just below $E_\mathrm{F}$ in the calculations), and the fully-occupied `M-shaped' states just below the Fermi level. 

The marked difference in the calculated electronic structure along the $\Gamma$-M$_x$ and $\Gamma$-M$_y$ directions demonstrates that the large anisotropy in the measured electronic structure along these directions can be fully explained on the basis of the orthorhombic crystal structure, without invoking an electronically driven nematicity as reported in a very recent study~\cite{horio_electron-driven_2019}. Moreover, our calculations indicate that the flattened `M-shaped' dispersion at the Brillouin zone centre, clearly evident in the experimental electronic structure in Fig.~\ref{fig1}(c), is the result of a band hybridisation between the top of a hole-like band and the bottom of an electron-like band, with a gap opening at the Fermi level. We will show below that this band hybridisation is in fact key to understanding much of the important physics of Ca$_3$Ru$_2$O$_7$. 

\begin{figure}
	\centering
	\includegraphics[width=\columnwidth]{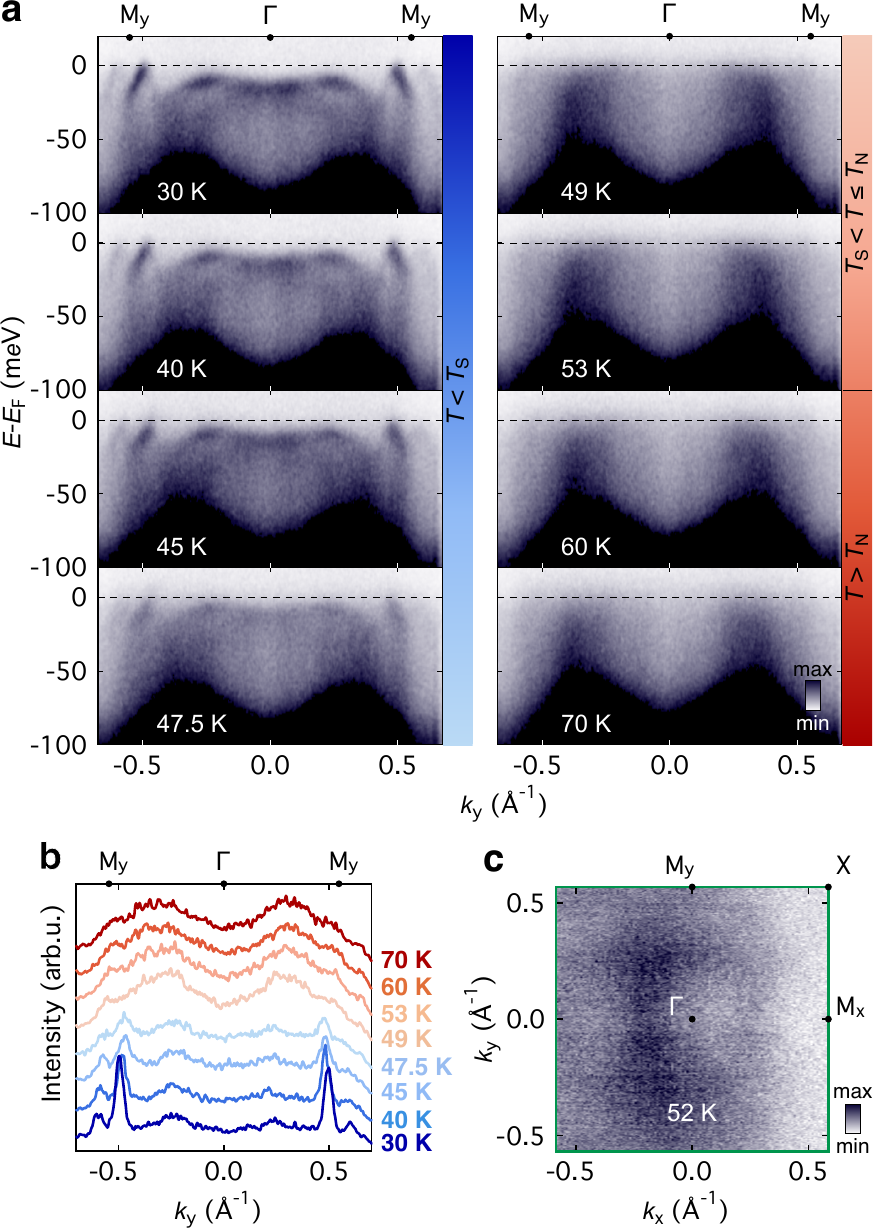}
	\caption{\textbf{Fermi surface transition at $\boldsymbol{T}_\mathbf{S}$.} (a) Temperature-dependent ARPES dispersions (measured along the $\Gamma$-M$_y$ direction) and (b) corresponding Fermi level momentum distribution curves (MDCs, $E_\mathrm{F}\pm2$~m{\em e}V) indicate a gradual temperature-dependent increase in the quasiparticle linewidth with increasing temperature below the coupled structural and spin-reorientation transition at $T_\mathrm{S}=48$~K. Upon warming through this transition, a sudden and extreme spectral broadening is observed, pointing to a loss of quasiparticle coherence at the transition. The `M-shaped' states at low temperature now disperse up through the Fermi level, forming (c) a large Fermi surface, but with a very high scattering rate. No qualitative changes are observed upon warming further through the N{\'e}el transition at $T_\mathrm{N}=56$~K (a).} 
	\label{fig2}
\end{figure}
Fig.~\ref{fig2}(a) shows the temperature-dependent evolution of the electronic structure through the two phase transitions. For temperatures below the structural and spin-reorientation transition, $T_\mathrm{S}$, the spectral linewidths gradually broaden with increasing sample temperature (Fig.~\ref{fig2}(b)), as can generically be expected from electron-electron and electron-phonon interactions. While Ca$_3$Ru$_2$O$_7$ is sometimes considered to be in an insulating state above $T\approx30$~K, our spectroscopic measurements clearly demonstrate that well-defined quasiparticle-like states persist up to $T_\mathrm{S}$. In contrast, we find a sudden and dramatic loss of quasiparticle coherence when warming through $T_\mathrm{S}$, with extremely broad linewidths above the transition indicative of a high scattering rate. The electronic structure is also markedly altered. While weak and broad remnants of the original Fermi crossings still persist (evident as shoulders at $k_y\approx\pm0.55$~\AA$^{-1}$ in Fig.~\ref{fig2}(b)), the state around the Brillouin zone centre now no longer appears to bend back to form an `M-shaped' dispersion. Rather, it crosses directly through $E_\mathrm{F}$, forming a large hole-like Fermi surface centred at $\Gamma$ (Fig.~\ref{fig2}(c)), consistent with a known transition in the Hall coefficient from large and negative values at low temperatures to small and positive values above $T_\mathrm{S}$\cite{yoshida_hall_2007,xing_existence_2018}. We note in passing that this high temperature Fermi surface is two-fold rather than four-fold symmetric, again reflecting the large orthorhombicity of the lattice.

We focus below on the origin of the unusual moment orientation-dependent Fermi surface transition at $T_\mathrm{S}$. Our measurements indicate that once the large zone-centred Fermi surface is established upon warming through $T_\mathrm{S}$, the electronic structure evolves only gradually, with no further qualitative changes as the temperature is increased to above the 56~K N{\'e}el transition (Fig.~\ref{fig2}(a,b)). Such an insensitivity to the antiferromagnetic ordering at $T_\mathrm{N}$ might naively suggest that it is the structural, rather than spin-reorientation, aspect of the phase transition at $T_\mathrm{S}$ which underpins the dramatic changes in electronic structure observed there. We show below, however, that this is in fact not the case. 

\begin{figure}[!h]
	\centering
	\includegraphics[width=\columnwidth]{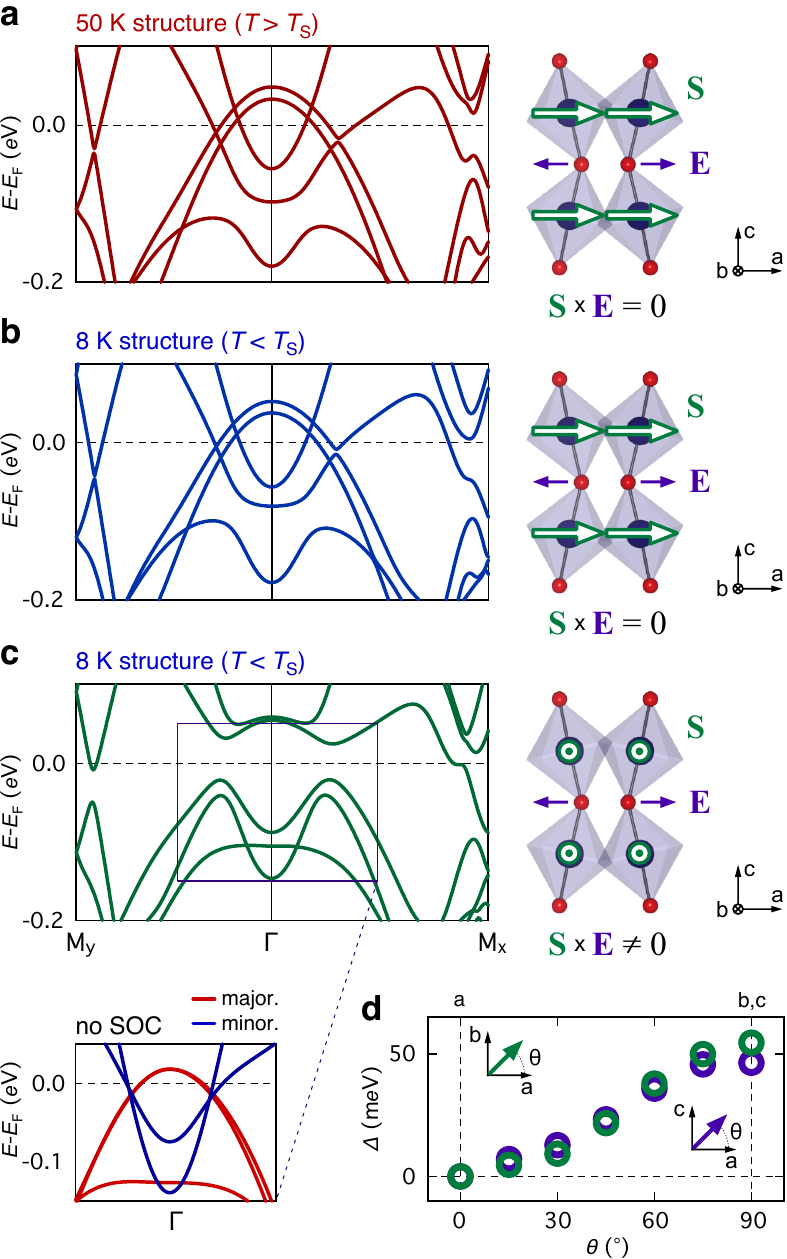}
	\caption{\textbf{Asymmetric spin-orbit driven band hybridisation.} (a) Calculated electronic structure from density-functional theory for the experimental crystal structure at $T=50$~K ($T>T_\mathrm{S}$) and with the spin moments aligned along the in-plane ${a}$-axis, as observed experimentally for $T_\mathrm{S}<T\leq{T_\mathrm{N}}$. (b,c) Equivalent calculations for (b) the low-temperature crystal structure (for $T=8$~K) with spin moments still oriented along the ${a}$-axis, and (c) the low-temperature crystal structure with spin moments along the ${b}$-axis, as observed experimentally for $T<T_\mathrm{S}$. The schematics in (a-c) show representative views of the Ru sites in a RuO$_2$ bilayer, showing the tilt and resulting local polarisation (purple arrows) and the spin moment orientation (green arrows). The inset of (c) shows the corresponding electronic structure calculated without including spin-orbit coupling, indicating that the hybridisation gap which opens at the Fermi level is between spin majority-like (red) and minority-like (blue) Ru $t_{2g}$ states, and opens via spin-orbit coupling. (d) The magnitude of the hybridisation gap grows smoothly as the moment is rotated away from the $a$ axis within both the $ab$ (purple points) and $ac$ (green points) plane. }
	\label{fig3}
\end{figure}
Fig.~\ref{fig3} shows the electronic structure as calculated by DFT for the experimental crystal structures above and below the structural transition at $T_\mathrm{S}$, and for the spin oriented along different in-plane crystallographic directions. For the spin moment oriented along the $a$-axis and for the 50~K crystal structure (Fig.~\ref{fig3}(a), as found for $T_\mathrm{S}<T\leq{T_\mathrm{N}}$), the hybridisation of the electron- and hole-like bands at the Brillouin zone centre is evidently suppressed. This gives rise to the large Fermi surfaces observed experimentally around the Brillouin zone centre (see Supplementary Fig.~S2 for a discussion of the relative band overlap and population of the electron vs.\ hole states {\it cf.} experiment). A similar lack of band hybridisation of the near-$E_\mathrm{F}$ states is found when considering the equivalent spin configuration but for the low-temperature crystal structure (Fig.~\ref{fig3}(b)), pointing to an insensitivity of the low-energy electronic structure to the structural component of the transition at $T_\mathrm{S}$. In contrast, when the moment is rotated to lie along the ${b}$-axis (or indeed $c$-axis; see Fig.~\ref{fig3}(d) and Supplementary Fig.~S3), the electron and hole bands at the zone centre develop a strong hybridisation. This demonstrates that it is the moment orientation and not the structural changes at $T_\mathrm{S}$ that mediate the opening of a band gap at the Fermi level observed here experimentally (Fig.~\ref{fig3}(c) and Fig.~\ref{fig1}(c)).

The relevant hybridised states at the Fermi level here derive from spin-minority-like and spin-majority-like Ru $t_{2g}$-derived bands (Fig.~\ref{fig3}(c, inset))~\cite{Note1}. Their hybridisation in the low-temperature phase is mediated by spin-orbit coupling (Fig.~\ref{fig3}(c)), with a gap that opens throughout the Brillouin zone (Fig.~\ref{fig1}(d) {\it cf.} Fig.~\ref{fig2}(c)) and develops gradually as the spin moment is rotated away from the $a$-axis (Fig.~\ref{fig3}(d)). We attribute this to a local breaking of inversion symmetry driven by the $X_3^-$ tilt mode of the RuO$_6$ octahedra~\cite{benedek_hybrid_2011,nowadnick_domains_2016}. The vertically-stacked octahedra of the perovskite bilayer develop hinge-like distortions about the shared apical oxygen, which reverse in direction between neighbouring in-plane sites as shown schematically in Fig.~\ref{fig3}. Each pair of outer apical oxygens are displaced in the opposite direction to the shared central apical oxygen. This leads to a local polarisation oriented along the tilt direction (the $a$-axis here) with an anti-ferro-type ordering within the $ab$ plane. No net polarisation is generated along $a$. Locally, however, an asymmetric spin-orbit coupling of the Rashba-type~\cite{bychkov_properties_1984,zhang_hidden_2014,riley_direct_2014}, $\mathcal{H}_{\mathrm{R}}\propto\mathbf{p}\cdot(\mathbf{S}\times\mathbf{E})$, can generically be expected, where $\mathbf{p}$ is the electron momentum, $\mathbf{E}$ is an effective internal electric field along the $a$-axis representing the local inversion asymmetry, and $\mathbf{S}$ is the electron spin of the itinerant states, which are fixed along one of the in-plane crystallographic axes by the magnetic moment orientation. This provides a natural explanation for the hybridisation of intra-bilayer spin-majority and minority bands observed here. For the spin moment aligned along the ${a}$-axis (Fig.~\ref{fig3}(a,b), $\mathbf{S}\times\mathbf{E}=0$), the Rashba-type term cannot act and thus no hybridisation would be expected from this form of spin-orbit coupling. In contrast, at the spin reorientation transition where the moment aligns along the ${b}$-axis (Fig.~\ref{fig3}(c)), the Rashba-like spin-orbit interaction becomes active ($\mathbf{S}\times\mathbf{E}\neq0$), enabling the band hybridisation.

\begin{figure}
	\centering
	\includegraphics[width=\columnwidth]{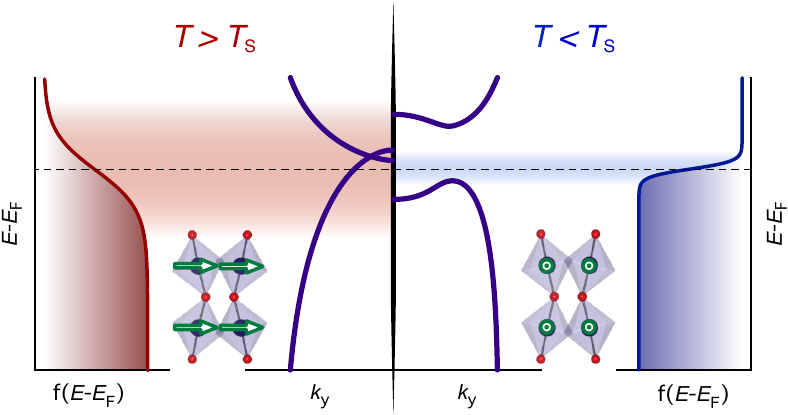}
	\caption{\textbf{Electronically driven magnetocrystalline anisotropy.} Schematic representation of the renormalised electronic structure of the zone-centre states corresponding to (left) $T>T_\mathrm{S}$ and (right) $T<T_\mathrm{S}$. At high temperatures, the broad Fermi function leads to thermal population of the states that are pushed upwards by band hybridisation in the low-temperature case, removing any associated energy gain with hybridising these states. Upon cooling, this energy gain becomes active, ultimately driving the spin reorientation in order to enable the band hybridisation via Rashba-type spin-orbit coupling. }
	\label{fig4}
\end{figure}
When the hybridisation is allowed, the large hole-like Fermi surface thus becomes gapped. The corresponding hybridisation energy scale in our DFT calculations is on the order of 50~m{\em e}V. In reality, however, the true hybridisation gap is renormalised to significantly smaller values due to the strong electronic correlations of this system. We estimate that the true gap magnitude in the low-temperature phase is $\approx10-15$~m{\em e}V, comparable to a 13~m{\em e}V gap (originally attributed as a pseudogap) which was seen to open in optical spectroscopy measurements upon cooling through $T_\mathrm{S}$~\cite{lee_pseudogap_2007}. The gap size is thus comparable to thermal energy scales at $T_\mathrm{S}$. We propose that at the N{\'e}el transition, the fluctuating moments of the paramagnetic state above $T_{\mathrm{N}}$ develop a long-range order, with the spin orientation fixed along the ${a}$-axis by conventional magnetocrystalline anisotropy effects. There is no electronic incentive for orienting the moment along ${b}$, as the hybridised states above $E_\mathrm{F}$ would be thermally populated (Fig.~\ref{fig4}); the crystalline terms thus dominate. Upon further cooling, however, the thermal population effects are reduced. The electronic energy gain from the opening of a band gap at the Fermi level thus becomes sufficient to favour the spin reorientation such that local Rashba-type spin-orbit coupling can hybridise the relevant electronic states. 

Our results thus suggest a Rashba-mediated band hybridisation of low-energy states ultimately leads to the first-order magnetic, and accompanying structural, phase transition in Ca$_3$Ru$_2$O$_7$. Our analysis has a number of important implications. It identifies a new source of magnetic anisotropy in metals, where the magnetic moment direction is set by an electronic energy gain from gapping much of the Fermi surface. In that sense it is a `magnetoelectronic' rather than a magnetocrystalline anisotropy. Materials where inversion symmetry is broken either within the host crystal structure, or by heterostructuring approaches~\cite{kim_polar_2016,bousquet_improper_2008,warusawithana_artificial_2003} and at surfaces and interfaces~\cite{sunko_maximal_2017,caviglia_tunable_2010}, will provide interesting playgrounds in which to seek to stabilise such magnetoelectronic anisotropy via asymmetric spin-orbit coupling as reported here. Other crystalline symmetries could in principle also be utilised, with the electronic energy gain being driven by spin moment reorientations that break the symmetries which otherwise protect specific band crossings in the vicinity of the Fermi level. Conversely, external switching of the moment orientation, for example via application of magnetic fields, would provide a novel route to control gapping of the Fermi surface states, and through this induce pronounced responses in the transport properties of multi-orbital correlated magnets such as Ca$_3$Ru$_2$O$_7$ studied here.


\

\section{Methods}

\noindent {\bf Single crystal growth:} Single crystals of Ca$_3$Ru$_2$O$_7$ were grown using a floating zone method in a mirror furnace (Canon Machinery, model SCI-MDH))~\cite{perry_systematic_2004}. The crystal growth was performed in an atmosphere of a mixture of Ar and O$_{2}$ ($\mathrm{Ar}:\mathrm{O}_{2}=85:15$). In general, antiphase domains can be expected, and are visible via contrast in polarised light optical microscopy (Supplementary Fig. S4). We used this to select samples which are single-domain over a scale of at least $500\times500$~$\mu$m$^2$. The mono-domain nature of our resulting samples is further evident in our measured Fermi surfaces, which show a clear two-fold symmetry with no signatures of rotated features coming from different domains.

\

\noindent {\bf Angle-resolved photoemission:} ARPES measurements were performing using the BLOCH beamline of the Max-IV synchrotron and the I05 beamline of Diamond Light Source. Measurements were performed using $p$-polarised 22~{\em e}V synchrotron light. Additional data measured using $s$-polarised light is shown in Supplementary Fig.~S1. The samples were cleaved {\it in situ}, and measured at temperatures between 6~K and 70~K, as specified in the figures. Temperature-dependent data sets were repeated on multiple samples and via both warming and re-cooling cycles, confirming that the changes shown in Fig.~\ref{fig2} are intrinsic and are not a result of sample ageing upon temperature cycling.

\

\noindent {\bf Density-functional theory:} Density functional theory (DFT) calculations were performed using the Local Spin Density Approximation (LSDA) exchange-correlation functionals, as implemented in the full-potential FPLO code~\cite{koepernik_full-potential_1999,opahle_full-potential_1999,noauthor_http://www.fplo._nodate}. Additional calculations were performed with the Perdew-Burke-Ernzerhof (PBE) functional~\cite{perdew_generalized_1996}, and are shown and discussed in Supplementary Fig.~S2. The experimental crystal structures were used in all cases~\cite{yoshida_crystal_2005}, and spin-orbit coupling was included throughout. The Brillouin zone sampling employed a $k$-mesh of at least $16\times 16\times 6$ $k$-points. Additional calculations were performed using WIEN2K~\cite{blaha_wien2k_nodate}, and gave consistent results. We employed ferromagnetic calculations, neglecting the antiferromagnetic coupling between neighbouring bilayers. Given the ferromagnetic ordering within the bilayer and the weak coupling between bilayers, this does not affect any of the key conclusions drawn from our calculations, as confirmed by the broad agreement between our calculations and the experimentally measured electronic structure shown in Fig.~\ref{fig1}(e).

\

\noindent {\bf Acknowledgments}
\noindent We thank Erez Berg, Bernd Braunecker, Sean Hartnoll, Phil Lightfoot, Finlay Morrison, Roderich Moessner, Silvia Picozzi, Ulrich R{\"o}ßler, Andreas Rost, and Veronika Sunko for useful discussions. We gratefully acknowledge support from the European Research Council (through the ERC-714193-QUESTDO project), the Royal Society, the UKRI (via grant number EP/R031924/1) the Max-Planck Society, and the JSPS KAKENHI (Nos. JP17H06136 and JP18K04715) and JST-Mirai Program (No. JPMJMI18A3) in Japan. IM and EAM acknowledge studentship support through the International Max-Planck Research School for the Chemistry and Physics of Quantum Materials. We thank Diamond Light Source and Max-IV synchrotrons for access to Beamlines I05 (Proposal Nos.~SI21986 and SI25040) and BLOCH (Proposal No.~20180399), respectively, that contributed to the results presented here.

\bibliographystyle{naturemag}

\begin{thebibliography}{10}
\expandafter\ifx\csname url\endcsname\relax
  \def\url#1{\texttt{#1}}\fi
\expandafter\ifx\csname urlprefix\endcsname\relax\def\urlprefix{URL }\fi
\providecommand{\bibinfo}[2]{#2}
\providecommand{\eprint}[2][]{\url{#2}}

\bibitem{rabe_modern_2007}
\bibinfo{author}{Rabe, K.~M.}, \bibinfo{author}{Dawber, M.},
  \bibinfo{author}{Lichtensteiger, C.}, \bibinfo{author}{Ahn, C.~H.} \&
  \bibinfo{author}{Triscone, J.-M.}
\newblock \bibinfo{title}{Modern {Physics} of {Ferroelectrics}: {Essential}
  {Background}}.
\newblock In \emph{\bibinfo{booktitle}{Physics of {Ferroelectrics}}}, vol.
  \bibinfo{volume}{105}, \bibinfo{pages}{1--30} (\bibinfo{publisher}{Springer
  Berlin Heidelberg}, \bibinfo{address}{Berlin, Heidelberg},
  \bibinfo{year}{2007}).

\bibitem{scott_applications_2007}
\bibinfo{author}{Scott, J.~F.}
\newblock \bibinfo{title}{Applications of {Modern} {Ferroelectrics}}.
\newblock \emph{\bibinfo{journal}{Science}} \textbf{\bibinfo{volume}{315}},
  \bibinfo{pages}{954--959} (\bibinfo{year}{2007}).

\bibitem{di_sante_electric_2013}
\bibinfo{author}{Di~Sante, D.}, \bibinfo{author}{Barone, P.},
  \bibinfo{author}{Bertacco, R.} \& \bibinfo{author}{Picozzi, S.}
\newblock \bibinfo{title}{Electric {Control} of the {Giant} {Rashba} {Effect}
  in {Bulk} {GeTe}}.
\newblock \emph{\bibinfo{journal}{Advanced Materials}}
  \textbf{\bibinfo{volume}{25}}, \bibinfo{pages}{509--513}
  (\bibinfo{year}{2013}).

\bibitem{picozzi_ferroelectric_2014}
\bibinfo{author}{Picozzi, S.}
\newblock \bibinfo{title}{Ferroelectric {Rashba} semiconductors as a novel
  class of multifunctional materials}.
\newblock \emph{\bibinfo{journal}{Frontiers in Physics}}
  \textbf{\bibinfo{volume}{2}} (\bibinfo{year}{2014}).

\bibitem{anderson_symmetry_1965}
\bibinfo{author}{Anderson, P.~W.} \& \bibinfo{author}{Blount, E.~I.}
\newblock \bibinfo{title}{Symmetry {Considerations} on {Martensitic}
  {Transformations}: "{Ferroelectric}" {Metals}?}
\newblock \emph{\bibinfo{journal}{Physical Review Letters}}
  \textbf{\bibinfo{volume}{14}}, \bibinfo{pages}{217--219}
  (\bibinfo{year}{1965}).

\bibitem{shi_ferroelectric-like_2013}
\bibinfo{author}{Shi, Y.} \emph{et~al.}
\newblock \bibinfo{title}{A ferroelectric-like structural transition in a
  metal}.
\newblock \emph{\bibinfo{journal}{Nature Materials}}
  \textbf{\bibinfo{volume}{12}}, \bibinfo{pages}{1024--1027}
  (\bibinfo{year}{2013}).

\bibitem{puggioni_designing_2014}
\bibinfo{author}{Puggioni, D.} \& \bibinfo{author}{Rondinelli, J.~M.}
\newblock \bibinfo{title}{Designing a robustly metallic noncenstrosymmetric
  ruthenate oxide with large thermopower anisotropy}.
\newblock \emph{\bibinfo{journal}{Nature Communications}}
  \textbf{\bibinfo{volume}{5}}, \bibinfo{pages}{3432} (\bibinfo{year}{2014}).

\bibitem{benedek_ferroelectric_2016}
\bibinfo{author}{Benedek, N.~A.} \& \bibinfo{author}{Birol, T.}
\newblock \bibinfo{title}{‘{Ferroelectric}’ metals reexamined: fundamental
  mechanisms and design considerations for new materials}.
\newblock \emph{\bibinfo{journal}{Journal of Materials Chemistry C}}
  \textbf{\bibinfo{volume}{4}}, \bibinfo{pages}{4000--4015}
  (\bibinfo{year}{2016}).

\bibitem{kim_polar_2016}
\bibinfo{author}{Kim, T.~H.} \emph{et~al.}
\newblock \bibinfo{title}{Polar metals by geometric design}.
\newblock \emph{\bibinfo{journal}{Nature}} \textbf{\bibinfo{volume}{533}},
  \bibinfo{pages}{68--72} (\bibinfo{year}{2016}).

\bibitem{laurita_evidence_2019}
\bibinfo{author}{Laurita, N.~J.} \emph{et~al.}
\newblock \bibinfo{title}{Evidence for the weakly coupled electron mechanism in
  an {Anderson}-{Blount} polar metal}.
\newblock \emph{\bibinfo{journal}{Nature Communications}}
  \textbf{\bibinfo{volume}{10}}, \bibinfo{pages}{3217} (\bibinfo{year}{2019}).

\bibitem{yoshida_crystal_2005}
\bibinfo{author}{Yoshida, Y.} \emph{et~al.}
\newblock \bibinfo{title}{Crystal and magnetic structure of
  Ca$_3$Ru$_2$O$_7$}.
\newblock \emph{\bibinfo{journal}{Phys. Rev. B}} \textbf{\bibinfo{volume}{72}},
  \bibinfo{pages}{054412} (\bibinfo{year}{2005}).

\bibitem{lei_observation_2018}
\bibinfo{author}{Lei, S.} \emph{et~al.}
\newblock \bibinfo{title}{Observation of {Quasi}-{Two}-{Dimensional} {Polar}
  {Domains} and {Ferroelastic} {Switching} in a {Metal}, Ca$_3$Ru$_2$O$_7$}.
\newblock \emph{\bibinfo{journal}{Nano Letters}} \textbf{\bibinfo{volume}{18}},
  \bibinfo{pages}{3088--3095} (\bibinfo{year}{2018}).

\bibitem{bousquet_improper_2008}
\bibinfo{author}{Bousquet, E.} \emph{et~al.}
\newblock \bibinfo{title}{Improper ferroelectricity in perovskite oxide
  artificial superlattices}.
\newblock \emph{\bibinfo{journal}{Nature}} \textbf{\bibinfo{volume}{452}},
  \bibinfo{pages}{732--736} (\bibinfo{year}{2008}).

\bibitem{benedek_hybrid_2011}
\bibinfo{author}{Benedek, N.~A.} \& \bibinfo{author}{Fennie, C.~J.}
\newblock \bibinfo{title}{Hybrid {Improper} {Ferroelectricity}: {A} {Mechanism}
  for {Controllable} {Polarization}-{Magnetization} {Coupling}}.
\newblock \emph{\bibinfo{journal}{Physical Review Letters}}
  \textbf{\bibinfo{volume}{106}}, \bibinfo{pages}{107204}
  (\bibinfo{year}{2011}).

\bibitem{nowadnick_domains_2016}
\bibinfo{author}{Nowadnick, E.~A.} \& \bibinfo{author}{Fennie, C.~J.}
\newblock \bibinfo{title}{Domains and ferroelectric switching pathways in Ca$_3$Ti$_2$O$_7$ from first principles}.
\newblock \emph{\bibinfo{journal}{Physical Review B}}
  \textbf{\bibinfo{volume}{94}}, \bibinfo{pages}{104105}
  (\bibinfo{year}{2016}).

\bibitem{yoshida_quasi-two-dimensional_2004}
\bibinfo{author}{Yoshida, Y.} \emph{et~al.}
\newblock \bibinfo{title}{Quasi-two-dimensional metallic ground state of {Ca$_3$}{Ru$_2$}{O$_7$}}.
\newblock \emph{\bibinfo{journal}{Physical Review B}}
  \textbf{\bibinfo{volume}{69}}, \bibinfo{pages}{220411}
  (\bibinfo{year}{2004}).

\bibitem{kikugawa_ca3ru2o7:_2010}
\bibinfo{author}{Kikugawa, N.}, \bibinfo{author}{Rost, A.~W.},
  \bibinfo{author}{Hicks, C.~W.}, \bibinfo{author}{Schofield, A.~J.} \&
  \bibinfo{author}{Mackenzie, A.~P.}
\newblock \bibinfo{title}{Ca$_3$Ru$_2$O$_7$: {Density} {Wave} {Formation} and {Quantum}
  {Oscillations} in the {Hall} {Resistivity}}.
\newblock \emph{\bibinfo{journal}{Journal of the Physical Society of Japan}}
  \textbf{\bibinfo{volume}{79}}, \bibinfo{pages}{024704}
  (\bibinfo{year}{2010}).

\bibitem{bohnenbuck_magnetic_2008}
\bibinfo{author}{Bohnenbuck, B.} \emph{et~al.}
\newblock \bibinfo{title}{Magnetic structure and orbital state of Ca$_3$Ru$_2$O$_7$ investigated by resonant x-ray diffraction}.
\newblock \emph{\bibinfo{journal}{Physical Review B}}
  \textbf{\bibinfo{volume}{77}}, \bibinfo{pages}{224412}
  (\bibinfo{year}{2008}).

\bibitem{bao_spin_2008}
\bibinfo{author}{Bao, W.}, \bibinfo{author}{Mao, Z.~Q.}, \bibinfo{author}{Qu,
  Z.} \& \bibinfo{author}{Lynn, J.~W.}
\newblock \bibinfo{title}{Spin {Valve} {Effect} and {Magnetoresistivity} in
  {Single} {Crystalline} {Ca$_3$}{Ru$_2$}{O$_7$}}.
\newblock \emph{\bibinfo{journal}{Physical Review Letters}}
  \textbf{\bibinfo{volume}{100}}, \bibinfo{pages}{247203}
  (\bibinfo{year}{2008}).

\bibitem{sokolov_metamagnetic_2019}
\bibinfo{author}{Sokolov, D.~A.} \emph{et~al.}
\newblock \bibinfo{title}{Metamagnetic texture in a polar antiferromagnet}.
\newblock \emph{\bibinfo{journal}{Nature Physics}}
  \textbf{\bibinfo{volume}{15}}, \bibinfo{pages}{671--677}
  (\bibinfo{year}{2019}).

\bibitem{nakatsuji_ca_1997}
\bibinfo{author}{Nakatsuji, S.}, \bibinfo{author}{Ikeda, S.-I.} \&
  \bibinfo{author}{Maeno, Y.}
\newblock \bibinfo{title}{Ca$_2$RuO$_4$: {New} {Mott} {Insulators} of {Layered}
  {Ruthenate}}.
\newblock \emph{\bibinfo{journal}{Journal of the Physical Society of Japan}}
  \textbf{\bibinfo{volume}{66}}, \bibinfo{pages}{1868--1871}
  (\bibinfo{year}{1997}).

\bibitem{baumberger_nested_2006}
\bibinfo{author}{Baumberger, F.} \emph{et~al.}
\newblock \bibinfo{title}{Nested {Fermi} {Surface} and {Electronic}
  {Instability} in {Ca}$_3${Ru}$_2${O}$_7$}.
\newblock \emph{\bibinfo{journal}{Physical Review Letters}}
  \textbf{\bibinfo{volume}{96}}, \bibinfo{pages}{107601}
  (\bibinfo{year}{2006}).

\bibitem{horio_electron-driven_2019}
\bibinfo{author}{Horio, M.} \emph{et~al.}
\newblock \bibinfo{title}{Electron-driven ${C}_2$-symmetric {Dirac}
  semimetal uncovered in {Ca}$_3${Ru}$_2${O}$_7$}.
\newblock \emph{\bibinfo{journal}{arXiv:1911.12163 [cond-mat]}}
  (\bibinfo{year}{2019}).

\bibitem{tamai_high-resolution_2019}
\bibinfo{author}{Tamai, A.} \emph{et~al.}
\newblock \bibinfo{title}{High-{Resolution} {Photoemission} on {Sr}$_2${RuO}$_4$
  {Reveals} {Correlation}-{Enhanced} {Effective} {Spin}-{Orbit} {Coupling} and
  {Dominantly} {Local} {Self}-{Energies}}.
\newblock \emph{\bibinfo{journal}{Physical Review X}}
  \textbf{\bibinfo{volume}{9}}, \bibinfo{pages}{021048} (\bibinfo{year}{2019}).

\bibitem{acharya_evening_2019}
\bibinfo{author}{Acharya, S.} \emph{et~al.}
\newblock \bibinfo{title}{Evening out the spin and charge parity to increase
  {T}$_c$ in unconventional superconductor {Sr}$_{2}${RuO}$_{4}$}.
\newblock \emph{\bibinfo{journal}{Communications Physics}}
  \textbf{\bibinfo{volume}{2}}, \bibinfo{pages}{163} (\bibinfo{year}{2019}).

\bibitem{yoshida_hall_2007}
\bibinfo{author}{Yoshida, Y.}, \bibinfo{author}{Ikeda, S.-I.} \&
  \bibinfo{author}{Shirakawa, N.}
\newblock \bibinfo{title}{Hall {Effect} in Ca$_3$Ru$_2$O$_7$}.
\newblock \emph{\bibinfo{journal}{Journal of the Physical Society of Japan}}
  \textbf{\bibinfo{volume}{76}}, \bibinfo{pages}{085002}
  (\bibinfo{year}{2007}).

\bibitem{xing_existence_2018}
\bibinfo{author}{Xing, H.} \emph{et~al.}
\newblock \bibinfo{title}{Existence of electron and hole pockets and partial
  gap opening in the correlated semimetal {Ca}$_3${Ru}$_2${O}$_7$}.
\newblock \emph{\bibinfo{journal}{Physical Review B}}
  \textbf{\bibinfo{volume}{97}}, \bibinfo{pages}{041113}
  (\bibinfo{year}{2018}).

\bibitem{Note1}
\bibinfo{note}{The relative spin orientation is reversed from bilayer to
  bilayer, due to the antiferromagnetic coupling between bilayers. Nonetheless,
  the inter-bilayer hopping is small, and so an effective ferromagnetic
  description is a good starting point for considering local interactions that
  can hybridise these states.}

\bibitem{bychkov_properties_1984}
\bibinfo{author}{Bychkov, Y.~A.} \& \bibinfo{author}{Rashba, E.~I.}
\newblock \bibinfo{title}{Properties of a 2d electron gas with lifted spectral
  degeneracy}.
\newblock \emph{\bibinfo{journal}{JETP Letters}} \textbf{\bibinfo{volume}{39}},
  \bibinfo{pages}{78--81} (\bibinfo{year}{1984}).

\bibitem{zhang_hidden_2014}
\bibinfo{author}{Zhang, X.}, \bibinfo{author}{Liu, Q.}, \bibinfo{author}{Luo,
  J.-W.}, \bibinfo{author}{Freeman, A.~J.} \& \bibinfo{author}{Zunger, A.}
\newblock \bibinfo{title}{Hidden spin polarization in inversion-symmetric bulk
  crystals}.
\newblock \emph{\bibinfo{journal}{Nature Physics}}
  \textbf{\bibinfo{volume}{10}}, \bibinfo{pages}{387--393}
  (\bibinfo{year}{2014}).

\bibitem{riley_direct_2014}
\bibinfo{author}{Riley, J.~M.} \emph{et~al.}
\newblock \bibinfo{title}{Direct observation of spin-polarized bulk bands in an
  inversion-symmetric semiconductor}.
\newblock \emph{\bibinfo{journal}{Nature Physics}}
  \textbf{\bibinfo{volume}{10}}, \bibinfo{pages}{835--839}
  (\bibinfo{year}{2014}).

\bibitem{lee_pseudogap_2007}
\bibinfo{author}{Lee, J.~S.} \emph{et~al.}
\newblock \bibinfo{title}{Pseudogap {Dependence} of the {Optical}
  {Conductivity} {Spectra} of {Ca}$_3${Ru}$_2${O}$_7$: {A} {Possible}
  {Contribution} of the {Orbital} {Flip} {Excitation}}.
\newblock \emph{\bibinfo{journal}{Physical Review Letters}}
  \textbf{\bibinfo{volume}{98}}, \bibinfo{pages}{097403}
  (\bibinfo{year}{2007}).

\bibitem{warusawithana_artificial_2003}
\bibinfo{author}{Warusawithana, M.~P.}, \bibinfo{author}{Colla, E.~V.},
  \bibinfo{author}{Eckstein, J.~N.} \& \bibinfo{author}{Weissman, M.~B.}
\newblock \bibinfo{title}{Artificial {Dielectric} {Superlattices} with {Broken}
  {Inversion} {Symmetry}}.
\newblock \emph{\bibinfo{journal}{Physical Review Letters}}
  \textbf{\bibinfo{volume}{90}}, \bibinfo{pages}{036802}
  (\bibinfo{year}{2003}).

\bibitem{sunko_maximal_2017}
\bibinfo{author}{Sunko, V.} \emph{et~al.}
\newblock \bibinfo{title}{Maximal {Rashba}-like spin splitting via
  kinetic-energy-coupled inversion-symmetry breaking}.
\newblock \emph{\bibinfo{journal}{Nature}} \textbf{\bibinfo{volume}{549}},
  \bibinfo{pages}{492--496} (\bibinfo{year}{2017}).

\bibitem{caviglia_tunable_2010}
\bibinfo{author}{Caviglia, A.~D.} \emph{et~al.}
\newblock \bibinfo{title}{Tunable {Rashba} {Spin}-{Orbit} {Interaction} at
  {Oxide} {Interfaces}}.
\newblock \emph{\bibinfo{journal}{Physical Review Letters}}
  \textbf{\bibinfo{volume}{104}}, \bibinfo{pages}{126803}
  (\bibinfo{year}{2010}).

\bibitem{perry_systematic_2004}
\bibinfo{author}{Perry, R.} \& \bibinfo{author}{Maeno, Y.}
\newblock \bibinfo{title}{Systematic approach to the growth of high-quality
  single crystals of {Sr}$_3$Ru$_2$O$_7$}.
\newblock \emph{\bibinfo{journal}{Journal of Crystal Growth}}
  \textbf{\bibinfo{volume}{271}}, \bibinfo{pages}{134--141}
  (\bibinfo{year}{2004}).

\bibitem{koepernik_full-potential_1999}
\bibinfo{author}{Koepernik, K.} \& \bibinfo{author}{Eschrig, H.}
\newblock \bibinfo{title}{Full-potential nonorthogonal local-orbital
  minimum-basis band-structure scheme}.
\newblock \emph{\bibinfo{journal}{Physical Review B}} \textbf{\bibinfo{volume}{59}},
  \bibinfo{pages}{1743--1757} (\bibinfo{year}{1999}).

\bibitem{opahle_full-potential_1999}
\bibinfo{author}{Opahle, I.}, \bibinfo{author}{Koepernik, K.} \&
  \bibinfo{author}{Eschrig, H.}
\newblock \bibinfo{title}{Full-potential band-structure calculation of iron
  pyrite}.
\newblock \emph{\bibinfo{journal}{Physical Review B}} \textbf{\bibinfo{volume}{60}},
  \bibinfo{pages}{14035--14041} (\bibinfo{year}{1999}).

\bibitem{noauthor_http://www.fplo._nodate}
\bibinfo{title}{http://www.fplo.de}.

\bibitem{perdew_generalized_1996}
\bibinfo{author}{Perdew, J.~P.}, \bibinfo{author}{Burke, K.} \&
  \bibinfo{author}{Ernzerhof, M.}
\newblock \bibinfo{title}{Generalized {Gradient} {Approximation} {Made}
  {Simple}}.
\newblock \emph{\bibinfo{journal}{Physical Review Letters}}
  \textbf{\bibinfo{volume}{77}}, \bibinfo{pages}{3865--3868}
  (\bibinfo{year}{1996}).

\bibitem{blaha_wien2k_nodate}
\bibinfo{author}{Blaha, P. e.~a.}
\newblock \bibinfo{title}{{WIEN}2k package, {Version} 10.1 (2010); available
  at, http://www.wien2k.at.}

\end{thebibliography}

\end{document}


\title{Supplementary Material: Electronically driven spin-reorientation transition of the correlated polar metal Ca$_3$Ru$_2$O$_7$}
	
\author{I.~Markovi{\'c}}
\affiliation {SUPA, School of Physics and Astronomy, University of St Andrews, St Andrews KY16 9SS, United Kingdom}
\affiliation{Max Planck Institute for Chemical Physics of Solids, N\"{o}thnitzer Strasse 40, 01187 Dresden, Germany}

\author{M.~D.~Watson}
\author{O.~J.~Clark}
\author{F.~Mazzola}
\affiliation {SUPA, School of Physics and Astronomy, University of St Andrews, St Andrews KY16 9SS, United Kingdom}

\author{E.~Abarca Morales}
\affiliation{Max Planck Institute for Chemical Physics of Solids, N\"{o}thnitzer Strasse 40, 01187 Dresden, Germany}
\affiliation {SUPA, School of Physics and Astronomy, University of St Andrews, St Andrews KY16 9SS, United Kingdom}

\author{C.~A.~Hooley}
\affiliation {SUPA, School of Physics and Astronomy, University of St Andrews, St Andrews KY16 9SS, United Kingdom}

\author{H.~Rosner}
\affiliation{Max Planck Institute for Chemical Physics of Solids, N\"{o}thnitzer Strasse 40, 01187 Dresden, Germany}

\author{C.~M.~Polley}
\author{T.~Balasubramanian}
\affiliation{MAX IV Laboratory, Lund University, P. O. Box 118, 221 00 Lund, Sweden}

\author{S.~Mukherjee}
\affiliation {Diamond Light Source, Harwell Campus, Didcot, OX11 0DE, United Kingdom}

\author{N.~Kikugawa}
\affiliation{National Institute for Materials Science, Tsukuba, Ibaraki 305-0003, Japan}

\author{D.~A.~Sokolov}
\affiliation{Max Planck Institute for Chemical Physics of Solids, N\"{o}thnitzer Strasse 40, 01187 Dresden, Germany}

\author{A.~P.~Mackenzie}
\affiliation{Max Planck Institute for Chemical Physics of Solids, N\"{o}thnitzer Strasse 40, 01187 Dresden, Germany}
\affiliation {SUPA, School of Physics and Astronomy, University of St Andrews, St Andrews KY16 9SS, United Kingdom}
	
\author{P.~D.~C.~King}
\email{philip.king@st-andrews.ac.uk}
\affiliation {SUPA, School of Physics and Astronomy, University of St Andrews, St Andrews KY16 9SS, United Kingdom}

\date{\today}
\maketitle

\

\begin{figure*}
	\centering
	\includegraphics[width=\textwidth]{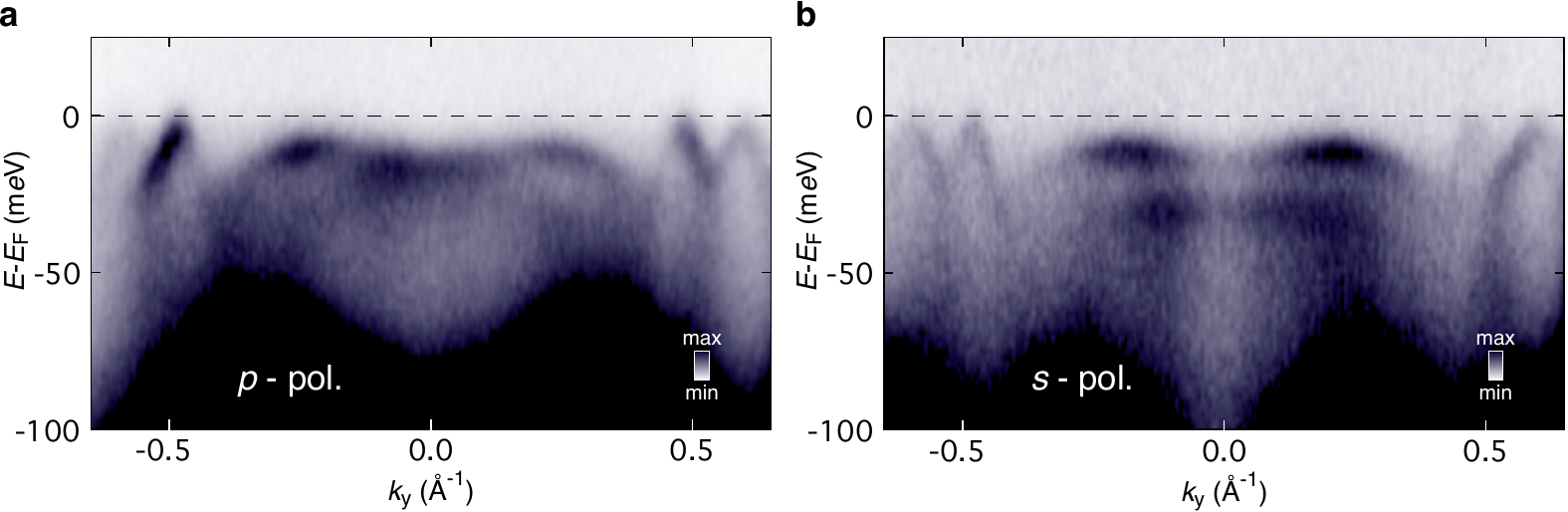}
	\caption{\textbf{Polarisation-dependent ARPES.} Experimental dispersions measured along the $\Gamma$-M$_y$ direction using (a) linear horizontal ($p$-) and (b) linear vertical ($s$-) polarised 22 $e$V synchrotron light at a sample temperature of 30 K. Due to the difference in the photoemission matrix elements with light polarisation, we can more clearly see both sides of the steep $\Lambda$-shaped bands close to the M$_y$ zone edge using $s$-polarised light, while the comparison between the two measurements clearly shows the multi-band structure of the flat bands at the zone centre.}
	\label{figS1}
\end{figure*}

\

\begin{figure*}
	\centering
	\includegraphics[width=0.9\textwidth]{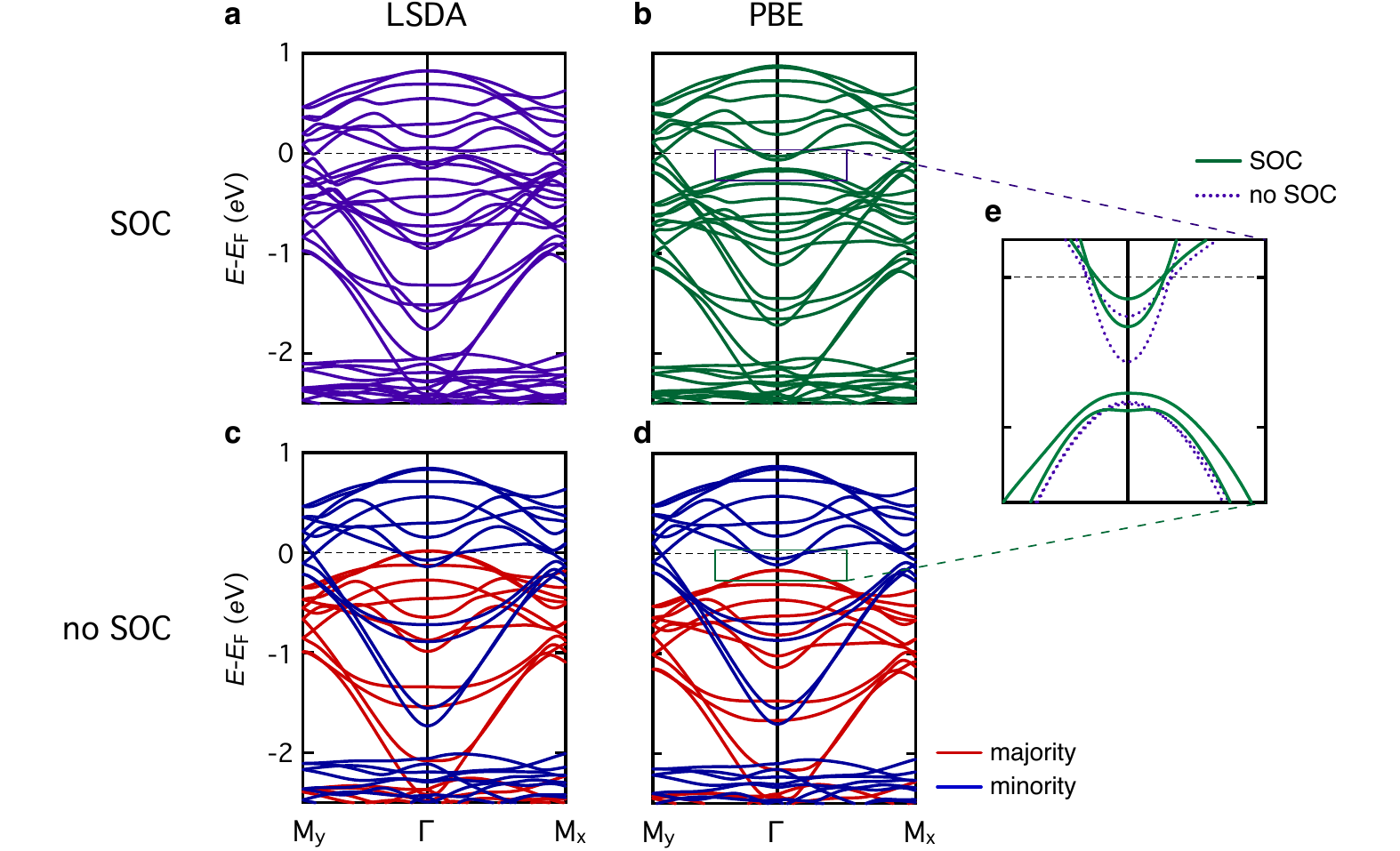}
	\caption{\textbf{Influence of exchange-correlation functional for DFT electronic structure.} Electronic structure along M$_y$-$\Gamma$-M$_x$ calculated assuming an FM-$b$ magnetic order (i.e. magnetic moment orientated along $b$) and the experimental crystal structure at 8 K\cite{Yoshida_Crystal_2005} using: (a,c) LSDA and (b,d) PBE functionals including (a,b) and excluding (c,d) spin-orbit coupling. As is often observed for generalised gradient (GGA) functionals vs. LDA ones, our PBE calculations exhibit a larger exchange splitting of the spin-majority (red) and spin-minority (blue) manifolds of Ru $t_{2g}$ states than for the LDA calculations. This leads to the hole-like bands at $\Gamma$ no longer intersecting the electron-like bands in the vicinity of the Fermi level. However, they are still close enough in energy such that spin-orbit coupling leads to a hybridisation, and consequent level repulsion, between the electron and hole bands for the FM-$b$ configuration (e), which is not present for the FM-$a$ or FM-$c$ phases. We expect the experimental electronic structure to lie  between the results for these two functionals, with the spin-majority hole-like and spin-minority electron-like bands close in energy, but potentially without significant band overlap. Our experimental measurements demonstrate that the hole bands at the zone centre in reality cross the Fermi level in the unhybridised state, and that the energy scale of the spin-orbit mediated band hybridisation is sufficient to push these states completely below the Fermi level. Since the experiments clearly show itinerant electrons at all temperatures, we refrain from using the Coulomb potential $U$ in our calculations, which would lead to a large separation of states at $\Gamma$~\cite{Liu_Mott_2011},  and a significant change in dispersion inconsistent with the experimentally observed "M-shaped" states at the zone centre at temperatures below $T_\mathrm{S}$. }
	\label{figS3}
\end{figure*}

\

\begin{figure*}
	\centering
	\includegraphics[width=\textwidth]{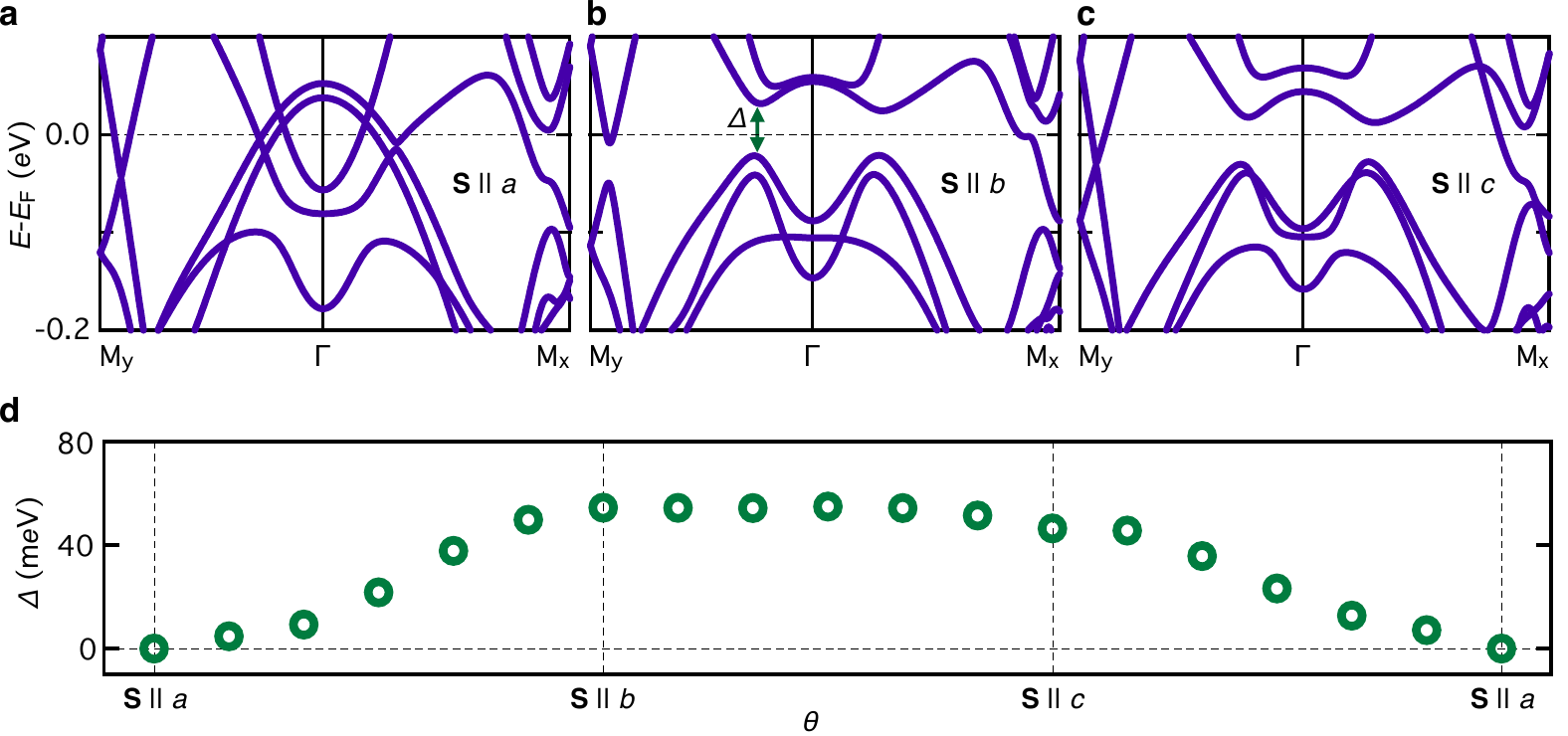}
	\caption{\textbf{Dependence of the Fermi surface hybridisation gap magnitude on magnetic moment orientation.} (a-c) Calculated low-energy electronic structures along $\Gamma$-M$_y$ and $\Gamma$-M$_x$ using LSDA+SOC calculations for the magnetic moment oriented along the (a) ${a}$, (b) ${b}$, and (c) ${c}$ crystal axes. (d) Magnitude of the hybridisation gap which opens at the Fermi crossing along $\Gamma$-M$_y$ as the magnetic moment orientation is rotated by angle $\theta$ between the ${a}$-${b}$-${c}$-${a}$ axes. The gap opens gradually as the magnetic moment rotates away from the ${a}$-axis, but is almost invariant when the moment is rotated between the $b$ and $c$ axes.}
	\label{figS4}
\end{figure*}

\
\newpage

\begin{figure*}
	\centering
	\includegraphics[width=\textwidth]{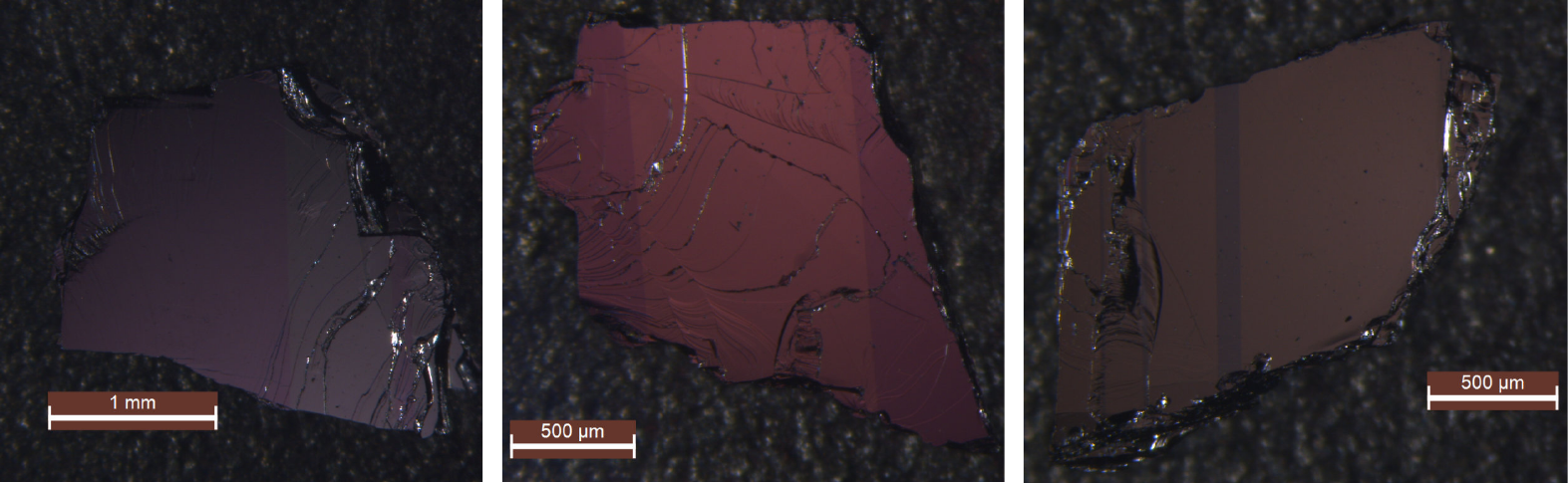}
	\caption{\textbf{Antiphase domains in Ca$_3$Ru$_2$O$_7$ single crystals.} To prepare the single crystals of Ca$_3$Ru$_2$O$_7$ for ARPES measurements, we used polarised light optical microscopy (example measurements shown here), where the change in contrast indicates the presence of antiphase domains, i.e. switching of the ${a}$ and ${b}$ (short and long) in-plane crystal axes. The crystals were then cut to $\approx\!500\times 500$ $\mu$m$^2$ size from a single antiphase domain, before being mounted for ARPES measurements and cleaved {\it in situ}.}
	\label{figS5}
\end{figure*}

\

\newpage

\

\

\

\bibliographystyle{naturemag}